%% file: 0_Main.tex
\documentclass[acmsmall,nonacm]{acmart}

\AtBeginDocument{%
  }

\usepackage{amsthm}
\usepackage{makecell}
\newtheorem{lemma}{Lemma}
\newtheorem{theorem}{Theorem}

\input{dataset_and_methods}

\usepackage[ruled,linesnumbered]{algorithm2e}
\begin{document}

\title[PPFedIT: Privacy-Preserving Federated Instruction Tuning]{\ourmethod: Towards Privacy-Preserving Federated Instruction
Tuning with Few-shot Local Examples}
\author{Zhuo Zhang}
\email{iezhuo17@gmail.com}
\affiliation{%
  \institution{Harbin Institute of Technology and Peng Cheng Lab}
  \city{Shenzhen}
  \country{China}
}
\author{Jingyuan Zhang}
\affiliation{%
  \institution{Kuaishou Technology}
  \city{Haidian, Beijing}
  \country{China}
}
\email{zhangjingyuan06@kuaishou.com}

\author{Jintao Huang}
\affiliation{%
 \institution{Harbin Institute of Technology}
 \city{Shenzhen}
 \country{China}}
\email{764695611@qq.com}

\author{Hui Wang}
\email{wanghu06@pcl.ac.cn}
\author{Yue Yu}
\affiliation{%
  \institution{Peng Cheng Lab}
  \city{Shenzhen}
  \country{China}
  }
\email{yuy@pcl.ac.cn}

\author{Hongzhi Zhang}
\affiliation{%
  \institution{Kuaishou Technology}
  \city{Haidian, Beijing}
  \country{China}
}
\email{zhanghongzhi@kuaishou.com}

\author{Xun Zhou}
\affiliation{%
 \institution{Harbin Institute of Technology}
 \city{Shenzhen}
 \country{China}
 }
\email{zhouxun2023@hit.edu.cn}
\author{Lizhen Qu}
\affiliation{
  \institution{Monash University}
  \city{Melbourne}
  \country{Australia}}
\email{Lizhen.Qu@monash.edu}
\author{Zenglin Xu}
\affiliation{%
 \institution{Fudan University}
 \city{Shanghai}
 \country{China}
}
\email{zenglinxu@fudan.edu.cn}
\renewcommand{\shortauthors}{Zhang et al.}

\begin{abstract}
Instruction tuning has been identified as a crucial technique for optimizing large language models (LLMs) to generate human-aligned responses. Nonetheless, gathering diversified and superior-quality instruction data for such tuning presents notable obstacles, especially in privacy-sensitive domains. Federated instruction tuning (\fedit) has emerged as a promising solution by consolidating collaborative training across multiple data owners, resulting in a privacy-enhancing learning model. Existing \fedit studies assume that clients have sufficient training data, however, in reality, many clients only have few-shot samples, leading to either overfitting in federated LLM or degraded performance. At the same time, this federated few-shot environment also increases the risk of training data extraction attacks, as the LLM may well memorize the limited training data. To address these issues, this paper proposes a \textit{novel} federated algorithm, \ourmethod, designed to enhance privacy protection and model performance of federated few-shot learning. \ourmethod comprises three vital steps on the client side: 
(1) synthetic data generation, which utilizes the strong generation capacity of LLMs to generate synthetic data, aiming to diversify and enrich local data; 
(2) parameter isolation training, which respectively updates the parameters of a shared global LLM on the synthetic data and the parameters of local LLMs on the local data, consequently mitigating the noise impact of the synthetic data; 
(3) local aggregation then sharing mechanism, which mixes the parameters of the global LLM and those of a local LLM first, before uploading them to a server for aggregation. This effectively mitigates data extraction attacks.
Extensive experiments on three open-source datasets demonstrate \ourmethod significantly enhances model performance (averaging 8.4\%) and reduces the risk of data extraction attacks (approximately 20\%) in practical and challenging federated few-shot scenarios.
\end{abstract}

\ccsdesc[500]{Computing methodologies~Natural language processing}
\ccsdesc[500]{Computing methodologies~Cooperation and coordination}

\keywords{Federated few-shot learning, Large Language Model, Data extraction attack.}

\setcopyright{cc}
\setcctype{by}
\copyrightyear{2026}
\acmYear{2026}
\acmDOI{10.1145/3806196}
\settopmatter{printacmref=false,printfolios=true}
\thanks{\copyright\ 2026 Copyright held by the owner/author(s).
This is the author's version of the work. The Version of Record was published in
\emph{ACM Transactions on Intelligent Systems and Technology} 17, 5 (2026), 1--23,
\url{https://doi.org/10.1145/3806196}.}

\maketitle

\section{Introduction}\label{sec:intro}
\input{1_Intro}

\section{Related Work}\label{sec:rwk}
\input{2_Related_Work}

\section{Method}\label{sec:meth}
\input{3_Method}

\section{Experiment}\label{sec:exp}
\input{4_Experiment}

\section{Conclusion}\label{sec:conc}
\input{5_Conclusion}

\section{Acknowledgments}
\input{6_acknowledgments}

\bibliographystyle{ACM-Reference-Format}
\bibliography{custom}

\input{7_Appendix}

\end{document}

%% file: dataset_and_methods.tex
\usepackage{xspace}

\newcommand{\methd}[1]{\texttt{#1}\xspace}
\newcommand{\data}[1]{\textsc{#1}\xspace}

\newcommand{\fedit}{\methd{FedIT}}
\newcommand{\ourmethod}{\methd{PPFedIT}}

\newcommand{\locit}{\methd{LocIT}}
\newcommand{\cenit}{\methd{Centralized}}
\newcommand{\fewfed}{\methd{FewFedWeight}}
\newcommand{\fedavg}{\methd{FedAvg}}
\newcommand{\fedprox}{\methd{FedProx}}
\newcommand{\scaffold}{\methd{SCAFFOLD}}
\newcommand{\fedopt}{\methd{FedOPT}}
\newcommand{\fedmit}{\methd{FedMIT}}

\newcommand{\alpaca}{\data{Alpaca}}
\newcommand{\alpacare}{\data{MedInstruct}}
\newcommand{\medalpaca}{\data{MedAlpaca}}

%% file: 1_Intro.tex
Instruction tuning is a crucial training step that allows large language models (LLMs) to understand user intentions and follow instructions directly from prompts~\cite{brown2020language,touvron2023llama,openai2023gpt}. Since instructions can vary by applications and users, collecting instruction-following data for training LLMs is time-consuming and labor-intensive. Despite the availability of public instruction-tuning datasets, it's estimated that high-quality public data will be depleted by 2026~\citep{villalobos2022will}.  Privacy concerns may further discourage users from sharing their data, such as private conversations and proprietary business data, especially when data sharing is strictly regularized by relevant legislation or regulations, such as the GDPR in EU, the HIPAA in the US, or PIPL in China.

To address the privacy concern, federated instruction tuning, coined \fedit, is proposed to leverage federated learning~\cite{konevcny2016federated,mcmahan2017communication} for collaboratively training instruction-following LLMs in a distributed environment~\cite{zhang2023towards,kuang2023federatedscope,fan2023fate,shen2023split,ye2024openfedllm}. As depicted in Fig.~\ref{fig: motivation} (a), \fedit exchanges model parameters instead of sharing private data among distributed data owners during instruction tuning, aiming to strike a promising balance between privacy protection and model performance.

Despite \fedit makes significant progress towards its goal, two significant challenges persist: 
(1) the existing \fedit algorithms assume each client has sufficient instruction data for training, which is difficult to satisfy in many real-world applications. Instead, local devices may merely hold a handful of data for demonstration, referred to as few-shot data, due to the high costs for data collection and annotation. In this case, traditional FL algorithms often experience significant performance degradation on these local devices~\cite{cai2023federated,cai2023towards}. As this setting is rarely explored in the prior studies on \fedit, it is desirable to address the limitations of \fedit algorithms for few-shot data. 
(2) It is reported that it is possible to extract training data from trained LLMs, known as \textit{training data extraction attack}~\cite{carlini2021extracting,brown2022does,nasr2023scalable}. As illustrated in Fig.~\ref{fig: motivation}, the attacker could leverage queries to extract private data from the models uploaded by clients. 
Although previous research on \fedit has been successful in mitigating model performance degradation on non-independent and identically distributed (non-IID) data~\cite{lin2021fednlp,zhang2023fedlegal} or reducing training costs~\cite{zhang2023fedpetuning, zhao2023fedprompt}, these studies have not investigated the risk of training data extraction attack. 
Consequently, we conduct preliminary experiments to uncover a \textit{new} key finding: \textit{the original \fedit with federated aggregation can mitigate privacy leakage in comparison to the training algorithms on centralized data, but it is still vulnerable to training data extraction attack, particularly in the federated few-shot learning setting.} 

To tackle the two challenges mentioned above, we propose a \textit{novel} federated instruction tuning algorithm, coined \ourmethod, to enhance both privacy preservation and model performance with few-shot examples in a practical and challenging federated environment. \ourmethod introduces three key steps on the client side, as illustrated in Fig. \ref{fig: motivation} (b). \textit{First}, to diversify and enrich local training data, we generate synthetic data on each local device by leveraging a global LLM shared across client devices in each federated round. Herein, this LLM employs the few-shot data as demonstration in each local device. 
\textit{Second}, to enhance the global LLM and filter out the noisy examples in the synthetic data, we employ a parameter isolation training strategy to train a reinforced local LLM as a filter for each local device by fine-tuning the global LLM on local data, and subsequently update the global LLM on the union of filtered synthetic data. 
The reinforced local LLMs are employed as: i) a quality measure to effectively select high-quality synthetic data, and ii) a way to mitigate noise in synthetic data that could otherwise degrade local LLM performance, which is different than directly mixed training~\cite{dong2022fewfedweight,cai2023towards}. Using this learning strategy, the global LLM captures privacy-agnostic knowledge shared across clients, while a local LLM preserves sensitive information of a client. \textit{Third}, to combine the complementary strengths of the global and local LLMs, we propose a local aggregation sharing mechanism, in which the client mixes the parameters of the local LLM with those of the global synthetic LLM before sharing them with the server. As a result, the parameters of the global model are aggregated from each local mixed model in each round. On the one hand, this local mixing mechanism facilitates knowledge fusion between the global LLM and the local LLMs. On the other hand, the local aggregation is analogous to "encrypting" the local LLM by using the global synthetic LLM as the public key. Our work contributions are outlined as follows:

\begin{figure}[t]
\centering
\includegraphics[trim={0cm 9cm 1.5cm 0cm}, clip, scale=0.43]{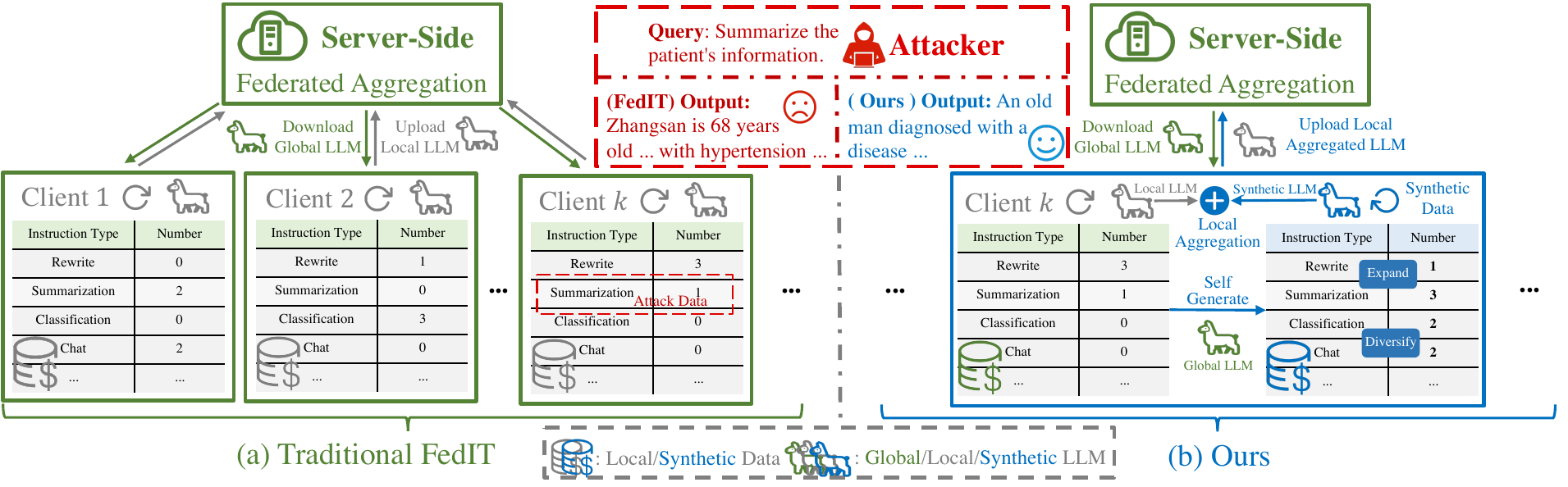}
\caption{(a) Overview of the traditional \fedit framework and (b) the workflow of \ourmethod (illustrated for client $k$). In contrast to \fedit, \ourmethod harnesses the generative power of the global LLM to expand and diversify the local data, boosting model utility in federated few-shot scenarios. Furthermore, to mitigate the notorious data extraction attacks that jeopardize the privacy of training data in \fedit, \ourmethod employs parameter-isolation training and local aggregation sharing, reinforcing client data safety throughout training.}
\label{fig: motivation} 
\end{figure} 

\begin{itemize}
    \item It is the first study to enhance \fedit performance and counter data extraction attacks within a real-world and challenging federated few-shot setting. We demonstrate that LLM trained in FL can effectively self-synthesize data, markedly improving model performance under the limited data scenario. Importantly, we discover that federated aggregation operation can mitigate data extraction attacks, leading us to propose a novel local aggregation sharing operation to utilize synthetic data to enhance privacy protection.

    \item We introduce a novel algorithm, termed \ourmethod, validated through extensive experiments across three datasets. The results demonstrate that our method improves model performance by an average of 6\% to 13\% and reduces privacy leakage by approximately 20\% compared to the baseline method. Further analysis indicates that our method effectively generates and filters high-quality instruction data, enriching the local dataset and thus narrowing the performance gap with centralized training.
    
\end{itemize}

%% file: 2_Related_Work.tex
\textbf{Federated instruction tuning.} Federated instruction tuning~\cite{zhang2023towards} offers a straightforward yet potent method for supporting distributed client privacy-preserving instruction tuning LLMs via the federated learning (FL) protocol \cite{konevcny2016federated, mcmahan2017communication}, bolstering LLMs' capabilities for handling privacy-sensitive real-world tasks. This research domain has garnered increasing attention. \citet{zhang2023towards} pioneered using FL for instruction tuning LLMs. Subsequently, various open benchmarks and repositories have facilitated research on federated instruction tuning tasks, including Federatedscope-llm~\cite{kuang2023federatedscope}, Fate-llm~\cite{fan2023fate}, OpenFedLLM~\cite{ye2024openfedllm}. However, these studies predominantly focus on constructing benchmarks for federated instruction tuning and do not propose advanced federation algorithms. Our work introduces a novel, more privacy-preserving federated instruction tuning algorithm, particularly adept in the federated few-shot setting. Recent work on federated NLP has emerged to enhance federated few-shot performance~\cite{cai2023towards,cai2023federated}. However, this research primarily focuses on classification tasks, utilizing pre-trained classification models and assuming extensive local unlabeled data availability. In contrast, we consider more complex instruction tuning tasks and synthetic data generation using the generative capabilities of LLMs.

\textbf{Training data extractable attack in language models.} Training data extraction attack \cite{carlini2021extracting,brown2022does,nasr2023scalable} has emerged as a tricky and unresolved challenge that illicit extraction of training data from LLMs. 
Such attacks typically exploit LLMs' memorization and prompt LLMs with some prefixes to generate training data.
\citet{nasr2023scalable} and \citet{carlini2021extracting} show larger or overfitted LLMs are more prone to leak training data.
In our work, we find that federated aggregation can diminish the risk of privacy data leakage compared to centralized training.
Thus, we introduce local aggregation sharing against training data extraction in the context of \fedit.

\textbf{Large language model as a data generator.} With LLMs revolutionizing the field of NLP, researchers have recently explored their potential as data generators~\cite{borisov2022language,wang2022self,zhang2023alpacare,li2023self,xu2023wizardlm,dubois2023alpacafarm,yehudai2024genie} to expand training datasets and reduce labor-intensive and expensive annotation costs. 
\citet{wang2022self} and \citet{zhang2023alpacare} curate high-quality, diverse seed data and upload it to AI systems (e.g., OpenAI) for data augmentation. 
However, this method is not feasible for privacy-preserving scenarios, as it prohibits private data uploading.
Alternatively, some research~\cite{li2023self} deploys sophisticated LLMs locally for on-device data augmentation.
Yet, advanced LLMs are typically proprietary, expensive, and computationally demanding.
In contrast, our approach leverages progressively enhanced federated LLMs as data generators to iteratively produce task-specific synthetic data throughout the federated process.
Recent research \cite{ye2023federated,huang2024federated} on FL with generative models has concentrated on using large models (e.g., diffusion models) to create additional data for addressing non-IID challenges. 
However, this research has primarily focused on image classification tasks, using image labels for data augmentation while neglecting the noise present in synthetic data. 
Compared to these studies, our work can self-generate synthetic data complemented by a filtering method to select high-quality synthetic data, achieving appealing performance on the complex federated instruction tuning task.

%% file: 3_Method.tex
In this section, we introduce the proposed \ourmethod for federated few-shot learning, a method of self-generating synthetic data for public-private parameter isolation training and privacy-enhancing aggregation sharing. 
We first provide the preliminaries of \fedit (Section~\ref{sec:3_1}), then overview \ourmethod (Section~\ref{sec:3_2}), and layout details about essential components of our framework: synthetic data generation (Section~\ref{sec:3_3}), parameter isolation training, and local aggregation sharing (Section ~\ref{sec:3_4}). 

\subsection{Preliminaries}\label{sec:3_1}
Suppose the standard federated setting comprises $N$ distributed clients with their local training data $\mathcal{D}_1, . . . , \mathcal{D}_N$ and a central server $\mathcal{S}$ responsible for coordinating global model training. 
Before training commences, $\mathcal{S}$ dispatches a global backbone model like LlaMa-2~\cite{touvron2023llama} to each client and determines global training parameters ($\mathcal{W}^g$) which are exchanged between the server and clients during each training round. Given the substantial communication and computational demands associated with the vast training parameters in LLMs, we adopt Low-Rank Adaption (LoRA)~\cite{hu2021lora} as all federated tuning strategies, aligning with prior research~\cite{zhang2023fedpetuning,zhang2023towards,fan2023fate,kuang2023federatedscope}. Thus, the $\mathcal{W}^g$ represents lightweight LoRA training parameters. 

In each round of federated training, the procedure alternates between \textit{server-side} and \textit{client-side} operations. In the client-side operation, each client uses local data to train the global model and then uploads updated parameters $\mathcal{W}^l$ to $\mathcal{S}$. During the server-side operation, $\mathcal{S}$ aggregates the client-upload model parameters to produce the updated global parameters for the next training round. This process is repeated multiple rounds until a certain condition is met (e.g., maximum communication rounds $\mathcal{R}$). 

\begin{figure}[t]
\centering
\includegraphics[trim={0cm 10.2cm 17.5cm 0cm}, clip, scale=0.7]{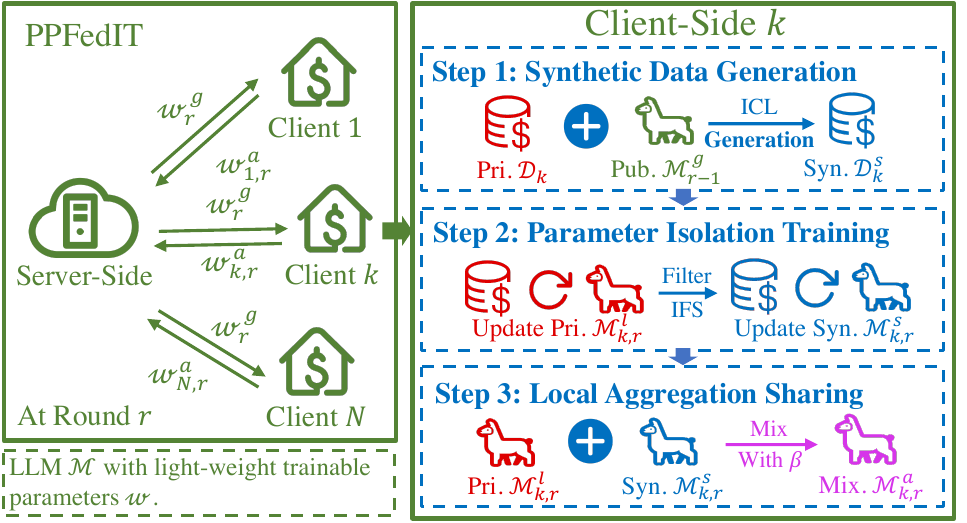}
\caption{
The overview of \ourmethod.
The innovation of \ourmethod compared to \fedit lies in client-side operations, including \textcircled{1} synthetic data generation, \textcircled{2} parameter isolation training, and \textcircled{3} local aggregation sharing.} 
\label{fig: head} 
\end{figure}

\subsection{Overview}\label{sec:3_2}
The innovation of \ourmethod lies in its client-side operations, which consist of three key steps: synthetic data generation, parameter isolation training, and local aggregation sharing.
We introduce private parameters $\mathcal{W}^l$, which are of the same size as the global parameters $\mathcal{W}^g$ and are managed locally by each client. In the synthetic data generation step, $\mathcal{W}^g$ self-generates data relevant to the local context. Subsequently, the client performs parameter isolation training, where $\mathcal{W}^l$ is trained on private local data $\mathcal{D}_k$, and $\mathcal{W}^s$ on high-quality synthetic data $\mathcal{D}_k^{s}$ filtered by $\mathcal{W}^l$. This step can further mitigate the impact of noise in synthetic data.
Inspired by experimental findings that federated aggregation can reduce privacy leakage, we implement local aggregation sharing, which combines $\mathcal{W}^s$ and $\mathcal{W}^l$ with a parameter $\beta$ before uploading to the server. This process reduces the exposure of privacy parameters, thereby mitigating the risk of local private data leakage. The comprehensive framework of our method is detailed in Algorithm \ref{algorithm}.

\subsection{Synthetic Data Generation}\label{sec:3_3}
Given the inherent data scarcity in federated few-shot scenarios, our approach leverages the concept of LLM-as-a-data-generator from previous work to augment local datasets~\cite{zhang2023alpacare,wang2022self,dubois2023alpacafarm}. 
However, the traditional LLM-as-a-data-generator approaches remain underutilized in privacy-sensitive contexts because they send private data to an open AI service provider (e.g., GPT-4), which violates data protection laws and incurs the high costs of numerous API calls.  
To circumvent these obstacles, we propose synthetic data generation to employ the federated instruction-tuned LLM as the data generator, complemented by a filtering method to ensure high-quality synthetic data. 
The underlying concept is that clients can enrich their local data by leveraging the global model that assimilates diverse instructions from multiple clients.
For convenience, we abbreviate the LLM with $\mathcal{W}^g$ as $\mathcal{M}^g$ and LLM with $\mathcal{W}^l$ as $\mathcal{M}^l$.
In the synthetic data generation step, the data generator $\mathcal{M}^g$ creates diverse candidate examples by using local data as demonstrations to exploit the in-context capabilities of LLM. Suppose $k$-th local client data $D_k$ contains $n_k$ triples $\{Instruction, Response\}$. The $\mathcal{M}^g$ generates synthetic candidate samples in two phases:

\textbf{Generate new reasonable instructions.} We randomly select eight instructions from local data as demonstrations to prompt $\mathcal{M}^g$ for new instruction generation. The Prompt for generating new instructions is provided in the Appendix~\ref{sec: app_prompts}. We then provide a format filter to eliminate failed instructions\footnote{For example, the failed generated sentences may begin with "As a ...", "sorry, ..." or be too short.}. We also follow prior research \cite{wang2022self,zhang2023alpacare} and use Rouge-L~\cite{lin2004rouge} similarity to amplify textual diversity. Namely, we discard new instructions with a Rouge-L similarity above 0.7 to any other local instructions. 
\textbf{Generate the corresponding responses given new instructions.}
We prompt $\mathcal{M}^g$ with randomly sampled examples followed by the new instruction (see the Prompt in Appendix~\ref{sec: app_prompts}). 
We require newly generated responses to conform to the template format (e.g., the generation contains the format prefixes $[Response]$). 
If the newly generated part does not meet the requirements, we will re-select demonstrations and generate them again. In our experiments, we discard the new instruction if it fails to generate a compliant sample three times. 
Additionally, we find the $\mathcal{M}^g$ may generate sentences directly copied from demonstrations due to weak instruction-following capability in early federated training. 
To mitigate this issue, we employ Rouge-L to filter out generated samples with over 0.7 similarity to any local examples. 
This also reduces the risk of leaking local data in synthetic data, as shown in Section~\ref{exp_sec: privacy}.
We utilize greedy decoding for all generations to encourage $\mathcal{M}^g$ to generate more grounded outputs~\cite{honovich2022unnatural}.

\input{algorithm}

\subsection{Parameter Isolation Training}
We rely on heuristic methods (e.g., Rouge-L or format) to filter out failed examples during synthetic data generation. However, whether the model generates accurate and contextually appropriate responses to new instructions remains uncertain, particularly in the early stages of federated training where the model may output low-quality responses due to weak generation capability. Using synthetic data of varying quality directly degrades model performance or even leads to performance collapse. It is critical to filter and utilize synthetic data effectively. Therefore, we propose a novel local training strategy, parameter-isolated training, to minimize the adverse effects of noisy synthetic data on model performance. 

Parameter isolation training begins by selecting appropriate training data from the generated examples. To conserve computational resources and protect potential privacy within the synthetic data, we refrain from using external models (such as GPT-4) as in prior research~\cite{wang2022self,du2023mods}. Instead, we use local privacy data to finetune an improved model $\mathcal{M}^l$ based on $\mathcal{M}^g$ and designate it as the evaluator to discern suitable examples from the generated candidates\footnote{Our preliminary experiments reveal that directly employing $\mathcal{M}^g$ to select synthetic data generated by itself result in overfitting and performance collapse.}.
Given the candidate example, we define the prompt $x$ as instruction with optional input and the output $y$ as the response. Typically, using the prompt $x$ as context simplifies the task for the LLM to generate $y$. We employ the \textit{Instruction Following Score} (IFS) to score each example and filter high-quality candidates~\cite{li2023quantity}.  
$\mathcal{M}^l$ calculates the conditional output loss $L(y|x)$ and the direct output loss $L(y)$, with the instruction following score defined as $IFS=\frac{L(y|x)}{L(y)}$. The IFS metric evaluates the degree of assistance the instruction provides in generating the corresponding response by comparing the model's response loss with and without instructional context. Thus, IFS can demonstrate how well the model aligns each response to the prompt. The low IFS indicates that the output $y$ closely aligns with the prompt $x$, signifying the example's quality. Finally, we sort the candidate examples by their IFS scores in ascending order and select the top $M$ examples as synthetic data. 

Despite rigorous synthetic data selection, noise may persist, particularly during the early stages of federated training due to the suboptimal performance of local models.
Parameter isolation training mitigates this problem by independently training the global model parameters $\mathcal{W}^{g}_{r-1}$ on filtered synthetic and local data. This process yields two distinct updated parameters: synthetic parameters $\mathcal{W}^{s}_{k,r}$, trained on synthetic data, and local parameters $\mathcal{W}^{l}_{k,r}$, trained on local data. 
Our method contrasts with traditional approaches that mix synthetic and private data and effectively reduce the influence of noisy synthetic data, as corroborated by experimental results in Section~\ref{exp_sec: privacy}. 
On the other hand, local parameters are learned from local training data, while synthetic parameters are derived from synthetic data. By separating these two types of parameters, we can more easily distinguish between sensitive local data and less critical synthetic data. This clear separation allows us to focus on protecting the local parameters, which are directly tied to private data, thereby enhancing privacy preservation. 

\subsection{Local Aggregation Sharing}\label{sec:3_4}
After parameter isolation training, we design the local aggregation sharing mechanism to improve model preference and resist data extraction attacks. 
Specifically, we use the hyperparameter $\beta$ to orchestrate the aggregation of local parameters and synthetic parameters: $\mathcal{W}_{k,r}^{a} = \beta*\mathcal{W}_{k,r}^{l} + (1-\beta) * \mathcal{W}_{k,r}^{s}$.
The hyperparameter $\beta$ controls the extent of exposure of the private parameters. 
High $\beta$ means the client exposes more privacy parameters and faces a higher risk of privacy leakage. Note that our approach degenerates into \fedit when $\beta=1$.
Consistent with federated aggregation, local aggregation integrates the knowledge from synthetic data and local data to enhance the model's performance in federated few-shot learning. 
Moreover, the update of synthetic data parameters $\mathcal{W}_{k,r}^s$ can act as "noise" injected into the locally private parameters $\mathcal{W}_{k,r}^l$, thereby strengthening the protection of client data. 
By prioritizing the protection of local parameters during aggregation and sharing processes, we can enhance the ability to protect privacy-sensitive data while still benefiting from the performance improvements offered by synthetic data.

%% file: algorithm.tex
\SetAlFnt{\small}
\SetAlCapFnt{\small}
\SetAlCapNameFnt{\small}
\begin{algorithm}
\DontPrintSemicolon
\SetKwInOut{Parameters}{Parameters}
\Parameters{

    Clients with their local data $\mathcal{C} = \{\mathcal{D}_1, \mathcal{D}_2, ..., \mathcal{D}_N\}$; Communication round $\mathcal{R}$; Randomly initialized global parameters $\mathcal{W}^{g}_0$ and local parameters $\mathcal{W}_{0}^{l,k}$ of $k$-th client; Self-generated and filtered synthetic data $\mathcal{D}_k^{s}$ of $k$-th client and its size $M$; Local aggregation sharing parameters ${\mathcal{W}_{k,r}^{a}}$ of $k$-th client; LocTrain refers to local training procedure; SynSel refers to synthetic data selection. \newline
}
\For{each communication round $r=1$ to $\mathcal{R}$} {
        $\mathcal{C}^{r} \leftarrow$ randomly sample K clients from $\mathcal{C}$\;
        Send global parameters $\mathcal{W}^{g}_{r-1}$ to selected clients\;
    	\For{each client $k \in \mathcal{C}^{r}$ \textbf{in parallel}}{
        \tcp{Synthetic Data Generation}
        $\mathcal{D}_k^{s} \leftarrow$ SynGen($\mathcal{D}_k$, $\mathcal{W}_{r-1}^{g}$)\;
        \tcp{Parameter Isolation Training}
        ${\mathcal{W}_{k,r}^{l}} \leftarrow $ LocTrain($\mathcal{D}_k$, $\mathcal{W}_{r-1}^{g}$)\;
        $\mathcal{D}_k^{s} \leftarrow$ SynSel($\mathcal{D}_k^{s}$, ${\mathcal{W}_{k,r}^{l}}$)\;
        ${\mathcal{W}_{k,r}^{s}} \leftarrow $ LocTrain($\mathcal{D}_k^{s}$, $\mathcal{W}_{r-1}^{g}$)\;
        \tcp{Local Aggregation sharing}
        ${\mathcal{W}_{k,r}^{a}} \leftarrow \beta*\mathcal{W}_{k,r}^{l} + (1-\beta) * \mathcal{W}_{k,r}^{s} $\;
        send ${\mathcal{W}_{k,r}^{a}}$ to the server
        }    	
        Perform federated aggregation on server:\;
        \quad \quad $\mathcal{W}_{r}^{g} \leftarrow \sum_{k=1}^{K} p_{k}\mathcal{W}_{k,r}^{a}$\;
        \quad \quad $p_k \leftarrow \frac{n_k}{\sum_{i=1}^K n_i}$\;
}
\caption{Training process of \ourmethod}
\label{algorithm}
\end{algorithm}

%% file: 4_Experiment.tex
This section showcases the effectiveness of \ourmethod through extensive experiments. 
We begin with the experiment setup in Section \ref{exp_sec: settings} and then report the performance assessment results in Section \ref{exp_sec: performance} and privacy defense evaluations in Section \ref{exp_sec: privacy}, respectively. 
Moreover, we provide further analysis in Section \ref{exp_sec: ablation} to explore how \ourmethod works. 

\subsection{Experimental Setup}\label{exp_sec: settings}
\input{results/datasets}
\textbf{Dataset and Partitions.} We evaluate the effectiveness of \ourmethod using three open-source instruction datasets, which contain one open domain \alpaca~\cite{peng2023instruction}, and two medical domains \alpacare~\cite{zhang2023alpacare} and \medalpaca~\cite{han2023medalpaca}. Previous work~\cite{li2023quantity,du2023mods,li2023one} demonstrated that the diversity and quality of data are essential in LLM instruction tuning. We follow \citet{li2023quantity} and use the KMeans algorithm to cluster each dataset into 100 clusters. Then, we take ten or five samples from each cluster to construct federated few-shot training settings for different domains.
The test sets for different datasets are the AlpacaEval~\cite{alpaca_eval} and MedInstructTest~\cite{zhang2023alpacare} for \alpaca and \alpacare, respectively. For \medalpaca, we randomly select 400 samples, ensuring no overlap between training and test sets. Please see Appendix~\ref{sec: app_datasets} for more dataset details. 
For the federated partition, we consider realistic and challenging non-IID data partitioning throughout the experiments~\cite{zhang2023fedpetuning}. 
In particular, we partition all datasets using the Dirichlet distribution~\cite{hsu2019measuring} with heterogeneity parameter 1.0 and cluster information as labels. 
Considering various FL scenarios, we set the amount of Cross-device clients to 50 and the amount of Cross-silo clients to 10. Table \ref{tab: data_fl} shows our experiment's data statistics and federated setting.

\textbf{Baselines.} Our experiment compares the proposed \ourmethod against the following baselines: \cenit is the skyline algorithm aggregating all data for model training without considering data privacy. \fedit represents the orthodox family of privacy-preserving instruction tuning algorithms, including \fedavg~\cite{mcmahan2017communication}, \fedprox~\cite{li2020federated}, \scaffold~\cite{karimireddy2020scaffold}, and \fedopt~\cite{reddi2020adaptive}. \fewfed~\cite{dong2022fewfedweight} advances federated few-shot learning by leveraging the global model to generate client pseudo labels and utilizing an energy-based algorithm to weight the pseudo samples.

\textbf{Evaluation Protocol.} Our experiment conducts free-form instruction evaluation for all methods using GPT-4\footnote{GPT-4-0613 version is used in our experiments.} as a judge. GPT-4 compares responses from an instruction-tuned LLM with reference responses from another LLM API (e.g., text-davinci-003 or GPT-3.5-turbo) or human for each corresponding instruction in the test sets. 
To improve the quality of the evaluations and mitigate the positional effects of GPT-4's assessments, we implement a dual-sided scoring system, as described in previous studies~\cite{zhang2023alpacare,zheng2024judging}. This system evaluates each output comparison twice, alternating the order of the instruction-tuned model output and the reference output. Appendix~\ref{sec: app_prompts} provides the evaluation prompt used by GPT-4. 

\input{results/utility}

\textbf{Models and Training Details.} Our experiment utilizes the LlaMa-2-7B model~\cite{touvron2023llama} because it is one of the most widely used large language models and is extensively employed in almost all \fedit studies. In real-world federated edge devices (e.g., smartphones and PCs), the LlaMa-2-7B model remains computationally demanding and has communication overhead. To address this, we employ LoRA~\cite{hu2021lora} with a low rank of 16 for training local LLMs across all baseline methods. 
Unless otherwise specified, we set the local epochs to 1 for all federated methods while setting the local epochs to 10 for \cenit. Following the cross-device protocol~\cite{ye2024openfedllm}, we randomly select two clients in each round. We use the AdamW optimizer with a training batch size of 8. \ourmethod generates 32 candidate samples in each self-generation step, filtering out 16 samples to create a synthetic dataset using IFS. By default, we set $\beta = 0.5$ as it provides the best trade-off between privacy loss and model utility. See Appendix~\ref{sec: app_datasets} for more training details.

\subsection{Utility Experiment}\label{exp_sec: performance}
Table \ref{tab: performace} presents the performance results of \ourmethod alongside baselines. We denote the WT score as the win and tie ratio sum. We can observe that \textbf{\ourmethod surpasses all federated algorithms with a substantial improvement 8.4\% on the average WT score and closely approaches the skyline algorithm \cenit performance (less than 3\%).} From Table~\ref{tab: performace}, \ourmethod exhibits a 7\% to 13\% improvement over the \fedit family within the federated few-shot context. Compared to the state-of-the-art \fewfed method, \ourmethod demonstrates a notable 6.3\% enhancement on the average WT score. While \fewfed also relies on pseudo-response generation during training to bolster federated few-shot performance, it fails to expand the diversity and scale of its local training data. In contrast, \textit{\ourmethod generates more useful task data, effectively mitigating data scarcity concerns.} We provide more synthetic data quality analysis in Appendix~\ref{app_quality}.

In Table \ref{tab: performace}, we observe that \fedit algorithms lag behind federated few-shot algorithms, while centralized algorithms consistently outperform all federated counterparts.
This observation underscores two key points: (1) Developing robust federated few-shot algorithms is essential, particularly when local clients possess limited training data.
(2) Non-IID data distribution remains a significant challenge for federated instruction tuning.
Notably, \ourmethod performs comparably to \cenit, demonstrating robustness against the non-IID challenge. This robustness is attributable to \textit{synthetic data generation, which integrates insights from other clients during the federated process, diversifying local data and enhancing overall training performance.}

\begin{figure}[t]
\centering
\includegraphics[trim={0cm 9.4cm 21.5cm 0cm}, clip, scale=0.6]{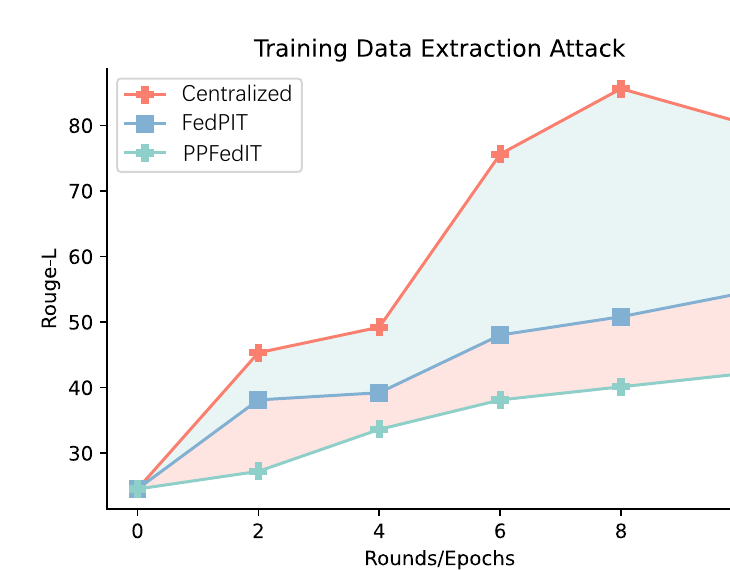}
\caption{
The privacy data leakage measurement of \cenit, \fedit and \ourmethod during the training process. The light green and light red areas reflect the data leakage gaps between \fedit and \cenit, as well as \fedit and \ourmethod, in the same training stage.}
\label{fig: privacy_gap} 
\end{figure} 

\begin{figure}[t]
\centering
\includegraphics[trim={3.6cm 0cm 2cm 0.3cm}, clip, width=\linewidth]{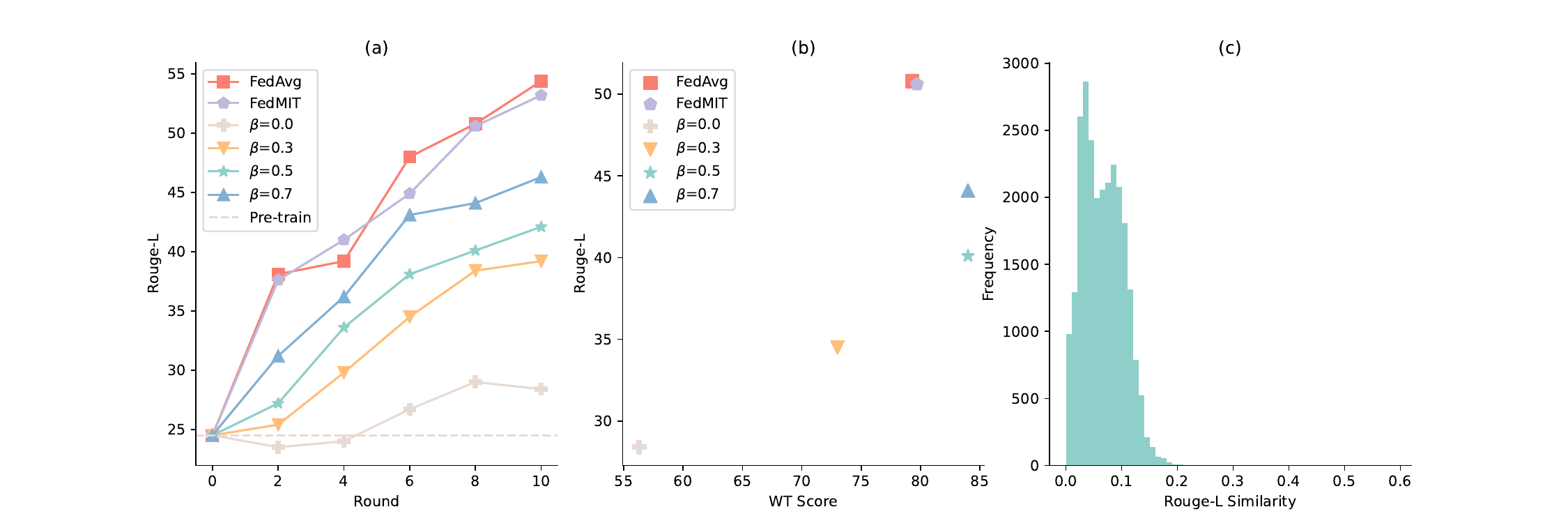}
\caption{
(a) shows privacy data leakage of \ourmethod with different $\beta$ and baselines proceeds with the federated training process. 
(b) shows the trade-off between privacy data leakage (\textit{y}-axis) and model utility (\textit{x}-axis). 
(c) shows the distribution of the Rouge-L similarity between private samples and their generated synthetic samples. 
We measure privacy data leakage using Rouge-L, where higher values indicate more significant privacy leakage. The model utility uses the WT score, representing the win and tie ratio sum. The parameter $\beta$ in \ourmethod determines how much client privacy parameters are exposed. Pre-train refers to the backbone model without instruction tuning. \ourmethod demonstrates stronger privacy-preserving capabilities against training data extraction attacks than all baselines.} 
\label{fig: privacy} 
\end{figure}

\subsection{Privacy Experiment}\label{exp_sec: privacy}
Next, we investigate the privacy-preserving capabilities of our method against the notorious training data extraction attack. 
We employ the discoverable memorization attack method elucidated by \citet{nasr2023scalable}.
Given a training string $s = [p||y] \in D_k$ that consists of a prefix $p$ and suffix $y$, the discoverable memorization attack prompts the threat model with the proper prefix $p$ to generate $y$. 
Discoverable memorization provides an upper bound estimation for the effectiveness of training data extraction attacks and is also prevalent in practical contexts. 
For example, attackers may target specific sensitive information such as "My bank card password is ..."~\cite{fowl2022decepticons} or clients may seek to gauge the extent of privacy leakage in their shared models.

\textbf{Setup.} Our privacy experiment considers an attacker who can access and query the threat model $\mathcal{M}_g$. This scenario is practical as the model is exposed during federated communication or deployed on the client side. 
We randomly select 100 examples from different clients in \alpacare to construct the attack dataset. 
For each example, the attacker exploits the prompt as the prefix $p$ and queries $\mathcal{M}_g$ to generate the response $y'$. 
We then evaluate the Rouge-L similarity between the actual $y$ and $y'$. The higher Rouge-L similarity indicates serious privacy leakage. 
Moreover, we investigate privacy leakage in synthetic samples since our method employs local privacy data as demonstrations for generating them. Specifically, we calculate the Rouge-L similarity between synthetic and private samples for each client.
We exploit the \fedavg and federated mixed training (\fedmit) as baselines. 
\fedmit refers to the direct integration of synthesized data and private data for local training. This approach does not involve parameter isolation training and local aggregation sharing operations compared to our method.
Since the $\beta$ parameter in our method regulates the exposure of private data during upload, we present experimental results for $\beta \in \{0.0, 0.3, 0.5, 0.7\}$. Note that our method degenerates to \fedavg when $\beta=1$. We also show the \texttt{Pre-train}, which refers to the backbone model without instruction tuning.

\textbf{Results.} Fig.~\ref{fig: privacy_gap} presents the privacy data leakage measurement of \cenit, \fedit and \ourmethod during the training process. We find the LLM under \cenit is more likely to memorize the training data than federated algorithms and leads to increased privacy leakage when the same data is used for training\footnote{In a cross-silo setting, the training data used in each round of \fedit is the same as that used in each epoch of \cenit.}. In contrast, \textbf{federated algorithms can mitigate these privacy leakage risks through federated aggregation operations.} However, regrettably, \fedit remains relatively vulnerable to data extraction attacks, highlighting the need for improved privacy preservation in FL frameworks.

Fig.~\ref{fig: privacy} presents the trade-off between privacy data leakage risk and model performance. 
We find \textbf{\ourmethod demonstrates stronger privacy-preserving capabilities against the training data extraction attack than all baselines.}
In Fig.~\ref{fig: privacy} (a), we observe that the risk of privacy data leakage increases with training duration across all methods. 
Among these methods, \fedavg($\beta=1.0$) exhibits the highest leakage, whereas \ourmethod with $\beta=0.0$ shows the lowest. 
\fedmit, which mixes synthetic and local data in its local training process, does not reduce privacy data leakage risk compared to \ourmethod and \fedavg. 
These results demonstrate that the data extraction attack in \fedit severely leaks clients' privacy-sensitive data, underscoring the necessity of our proposed local aggregation sharing operations.

As illustrated in Fig.~\ref{fig: privacy} (b), we find \ourmethod can flexibly adjust $\beta$ to achieve a better balance.
\fedmit produces results comparable to \fedavg but falls short of our method with $\beta=\{0.5,0.7\}$. These findings suggest that even carefully curated synthetic data also remains noise and parameter isolation training can effectively mitigate and enhance model performance. 
Surprisingly, our method with $\beta=0.0$ also achieves a 10.3\% win ratio, solely using synthetic data for federated aggregation. This result suggests that our method can generate useful data for improving model performance.

Fig.~\ref{fig: privacy} (c) plots the distribution of Rouge-L similarity scores between private samples and their generated synthetic ones.
We find that there is a significant disparity between synthetic data and local private data.
Our approach explicitly limits the overlap between synthetic and original data in Section~\ref{sec:3_3} to ensure that \textbf{the synthetic data does not leak information from local data.} This precaution is essential for further preserving client privacy. 

\begin{figure}[t]
\centering
\includegraphics[trim={2cm 0cm 1cm 0cm}, clip, scale=0.55]{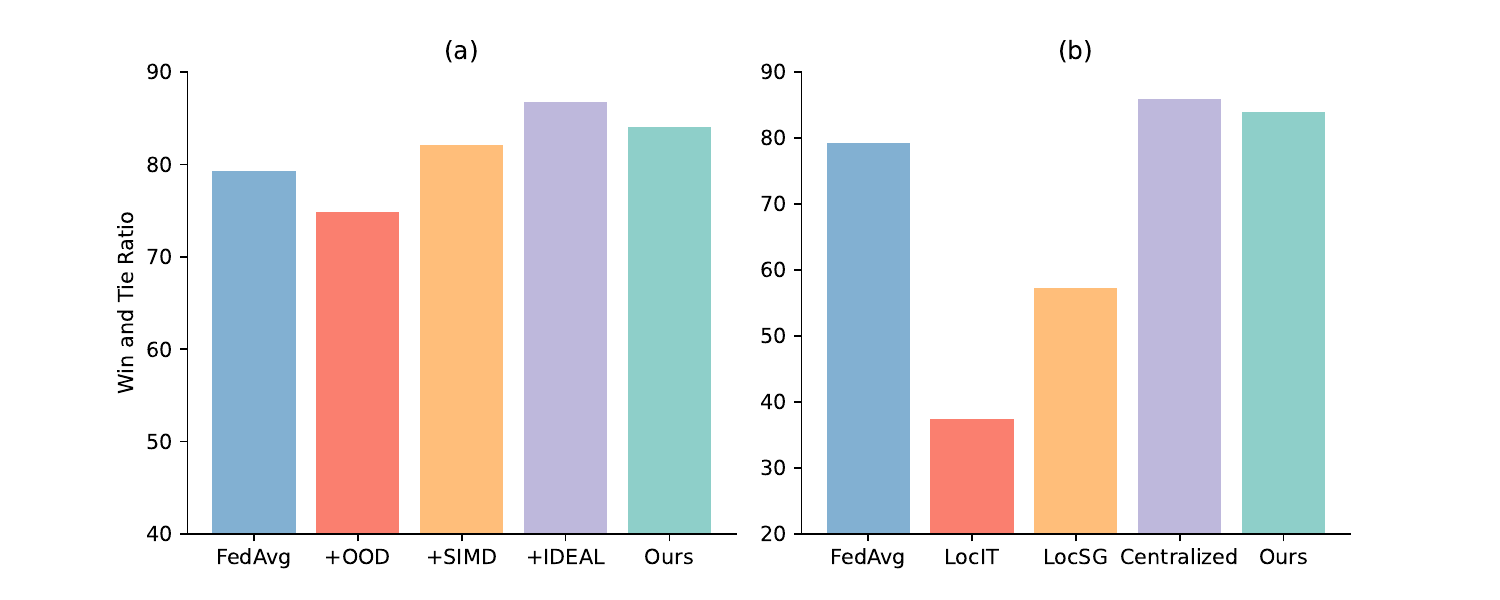}
\caption{(a) The WT scores of various replaced synthetic data during the federated training process. (b) The contribution of FL to synthetic data generation.} 
\label{fig: connect} 
\vspace{-0.4cm}
\end{figure}

\subsection{Further Analysis}\label{exp_sec: ablation}
To explore how \ourmethod works, we undertake a thorough analysis, including the contribution of local data and FL to synthetic data generation, the impact of different high-quality synthetic data selections, and the additional computational overhead.

\subsubsection{Contribution of Local Data to Self-generation.}
Within \ourmethod, synthetic data generation leverages local data as demonstrations to generate task-related data. 
As depicted in Table \ref{tab: performace}, integrating synthetic data enhances model performance in federated few-shot scenarios. 
We investigate whether injecting out-of-domain or domain-similar (e.g., medical instruction data) public data can also yield similar enhancements. 
We substitute the self-generated synthetic data with equal proportions of out-of-domain or domain-similar public data during federated training. 
Specifically, we employ \alpaca as the out-of-domain public data (+OOD) and \medalpaca for domain-similar public data (+SIMD). 
We randomly select the remaining training data from \alpacare (+IDEAL) as the ideal synthetic training data. 

Fig.~\ref{fig: connect} (a) presents the performances of various synthetic data in federated few-shot instruction tuning. 
The incorporation of domain-similar data (+SIMD and +IDEAL) enhances training performance, while the inclusion of out-of-domain data (+OOD) results in a decline (4.5\% performance gap compared to \fedavg). 
This result shows \textbf{adding synthetic data that is more similar to local data can improve federated training performance.} 
Our method exhibits a 2.1\% performance enhancement compared to +SIMD. 
These results highlight the significance of local data in self-generation.
Additionally, our method's performance closely approximates that of +IDEAL (less than 2\% WT score), indicating the high fidelity of our self-generated synthetic data.

\subsubsection{Can \ourmethod Discard FL?} Synthetic data generation shows the effectiveness of local dataset expansion. This raises the question of whether we can solely rely on local data without resorting to FL. To address this question, we conduct experiments implementing local instruction tuning (\locit) and local self-generation without FL (\locit+SG) on \alpacare. The outcomes in Fig.~\ref{fig: connect} (b) reveal that while \locit+SG significantly enhances local instruction tuning performance, the synthesized data remains inadequate and lags behind \fedavg and \ourmethod. This result underscores that \textbf{FL is still indispensable for data-scarce and privacy-sensitive downstream tasks.} 

\begin{figure}[t]
\centering
\includegraphics[trim={2cm 0cm 2cm 0.5cm}, clip, scale=0.55]{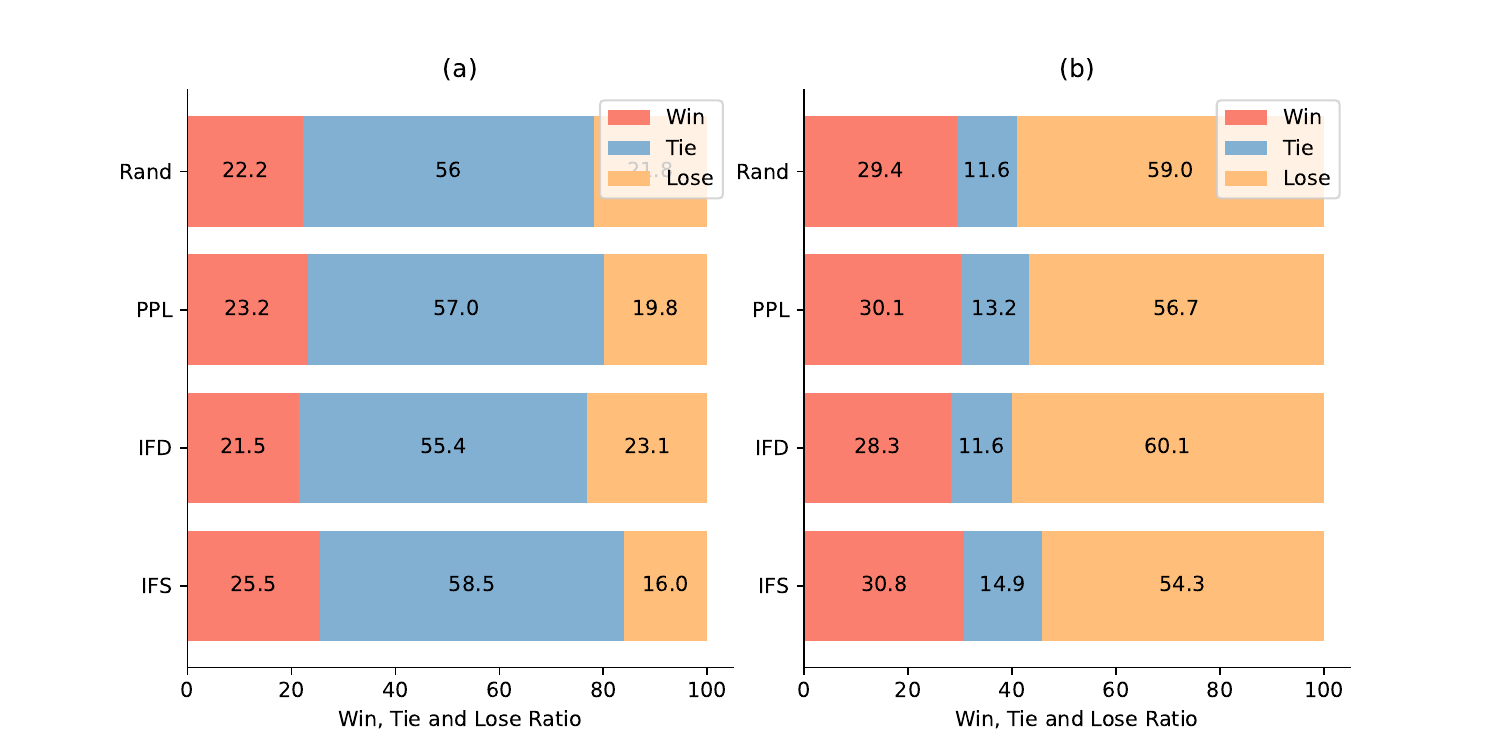}
\caption{The comparison of different synthetic data selection methods on \alpacare (a) and \medalpaca (b). Our IFS method more effectively screens higher-quality samples than the other methods.}\label{fig: selection}
\vspace{-0.4cm}
\end{figure}

\subsubsection{Performance under different Non-IID Settings}
Non-independent and identically distributed data represents a major challenge in federated learning, often leading to significant degradation in model performance. Following the configuration of \citet{lin2021fednlp}, we employ a Dirichlet distribution as the class prior to partition datasets. Specifically, we sample from $\mathcal{D} \sim Dir(\alpha)$ and assign the resulting subset $\mathcal{D}_k$ to client $k$. The concentration parameter $\alpha$ controls the degree of Non-IID: smaller values of $\alpha$ induce greater label distribution skew across clients.

\begin{figure}[t]
\centering
\includegraphics[trim={0cm 0cm 0cm 0cm}, clip, scale=0.45]{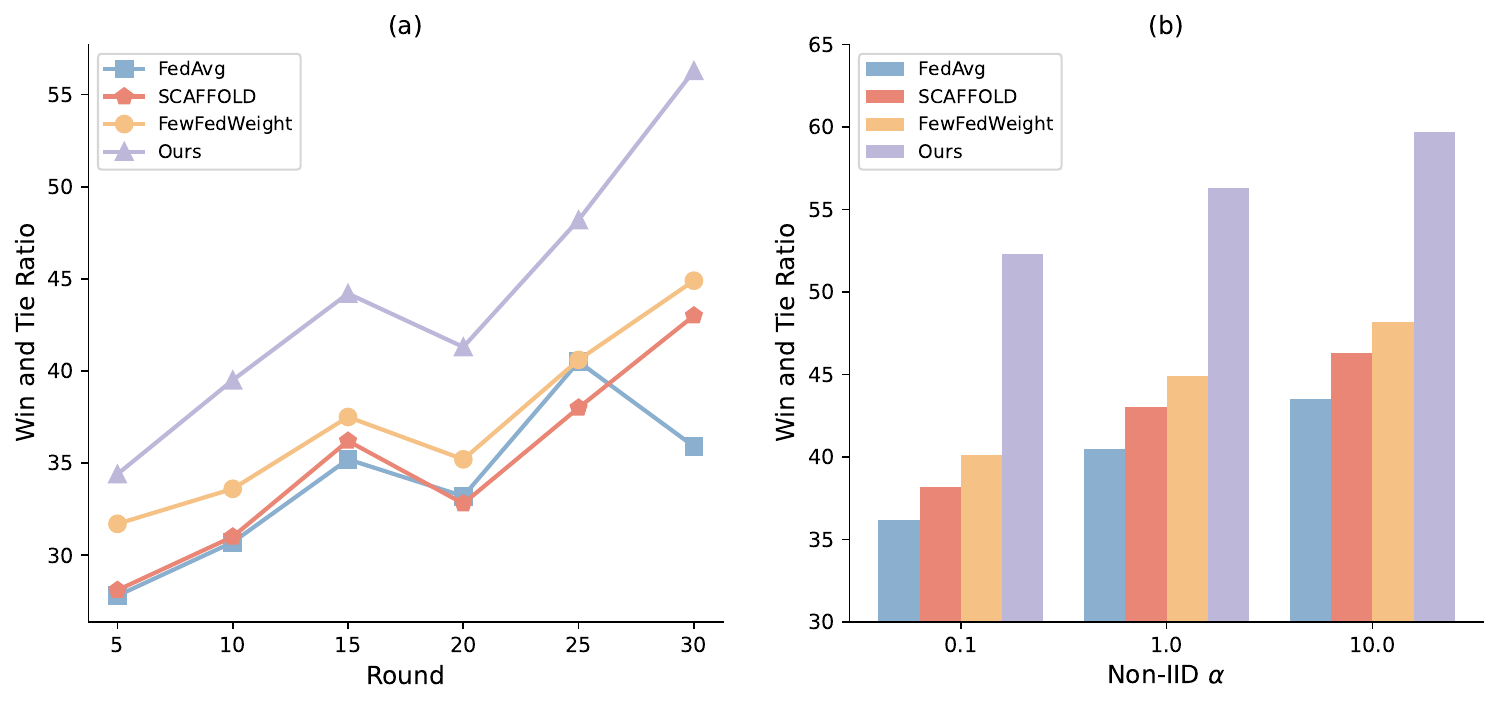}
\caption{(a) Convergence analysis of our method and the baseline methods on \alpaca and (b) Performances under different Non-IID Settings. 
}\label{fig: non-iid}
\vspace{-0.4cm}
\end{figure}

Fig.~\ref{fig: non-iid} (a) presents convergence analyses of various methods under the setting $\alpha=1$. We observe that our approach consistently outperforms competing baselines, achieving faster and more stable convergence to superior performance. We further examine the behavior of our method under different non-IID distributions. As shown in Fig.~\ref{fig: non-iid} (b), across a range of non-IID settings, our approach demonstrates substantial improvements over alternatives. Compared with the naïve \fedavg algorithm, \scaffold exhibits improved robustness due to its use of control variates to mitigate “client drift” in local updates. On the contrary, our method leverages globally shared models to generate synthetic data as a form of shared knowledge, thereby enhancing resilience to severe Non-IID settings.

\subsubsection{Impact of Different Synthetic Data Selection Methods.} Our experiment utilizes the IFS selection method to assess the alignment of model responses with corresponding prompts. 
We also investigate other scoring functions, including sequence perplexity (PPL), instruction following difficulty (IFD) \cite{li2023quantity}, and random selection (Rand). 
In Fig.~\ref{fig: selection}, we find \textbf{IFS can more effectively screen higher-quality samples than the other methods.} 
IFD selects the highest conditional probability and performs the worst. 
This finding highlights that prioritizing high-quality instructions is more crucial than selecting difficult ones, especially when dealing with noisy synthetic data.

\begin{figure}[t]
\centering
\includegraphics[trim={0cm 0cm 0cm 0cm}, clip, scale=0.45]{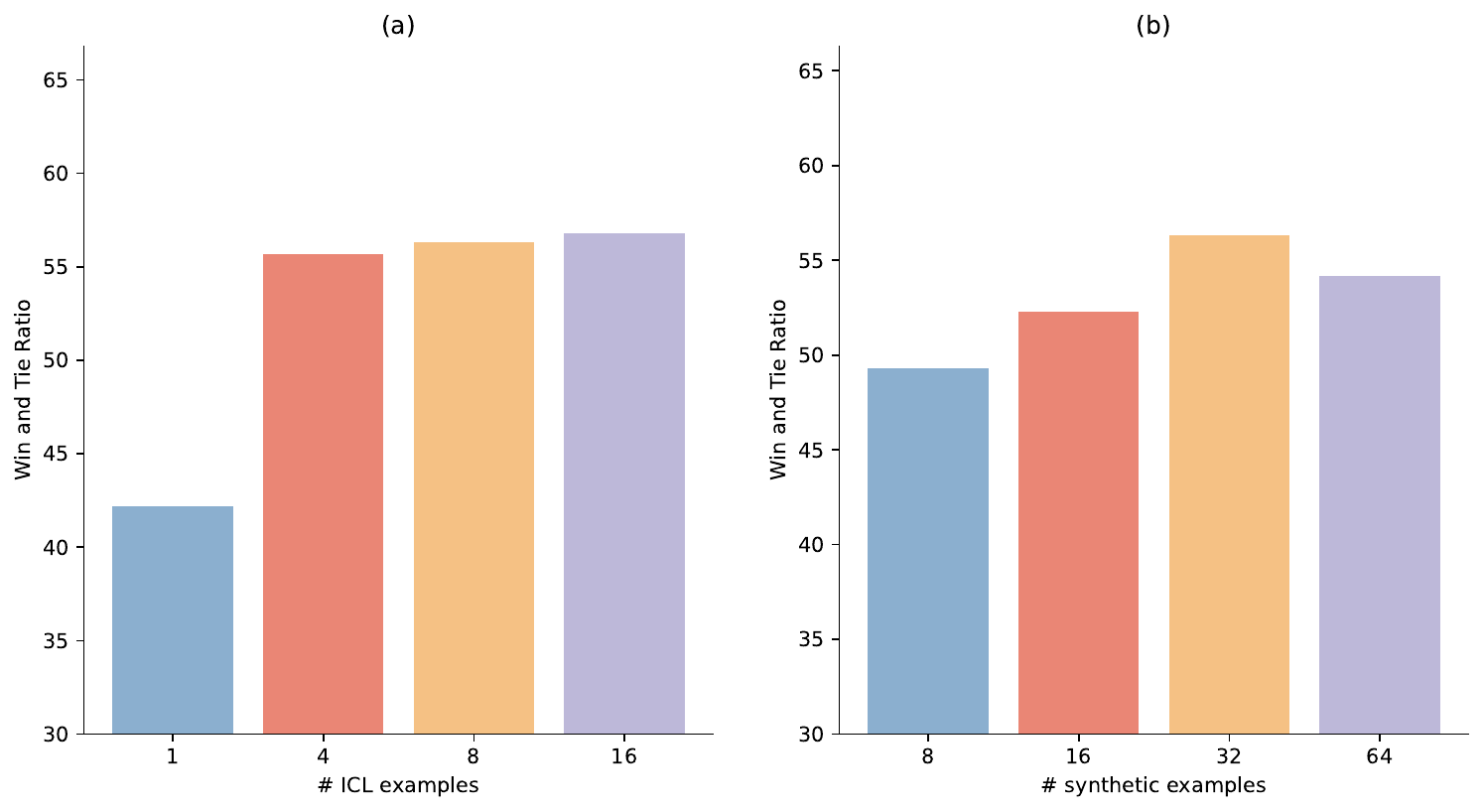}
\caption{Impact of demonstration sample size (a) and synthetic data volume (b) during synthetic data generation.
}\label{fig: synthetic_gen}
\vspace{-0.4cm}
\end{figure}

\subsubsection{Impact of demonstration size and synthetic data volume}
We systematically examine the effects of demonstration size and synthetic data volume on model performance. As shown in Figure~\ref{fig: synthetic_gen} (a), performance steadily improves with an increasing number of demonstration samples, underscoring the importance of richer contextual information. However, when the demonstration is severely limited (e.g., only a single sample), model performance deteriorates sharply, indicating insufficient guidance for effective generalization.
Figure~\ref{fig: synthetic_gen} (b) depicts the influence of synthetic data volume. While enlarging the synthetic dataset generally facilitates performance gains, excessive amounts can introduce substantial noise, thereby offsetting the benefits and even degrading performance. Furthermore, larger synthetic configurations entail additional inference and computational costs, raising practical concerns for deployment. In our experiments, we adopt 32 high-quality synthetic samples as the default training configuration, as this choice offers a favorable balance between efficiency and effectiveness.

\subsubsection{Additional computational overhead.} Incorporating synthetic data generation into \fedit enhances model performance but introduces additional inference computation and time overhead. Here, we provide additional computational inference time. 
When evaluating the \alpacare dataset with an inference batch size of 16, compared to the standard \fedit, our method incurs an average additional time cost of 2.9 minutes per round. Compared with the advanced \fewfed, \ourmethod introduces an average overhead of 0.6 minutes per round. 
The added burden is manageable since inference is lightweight compared to training and \ourmethod generates minimal synthetic data in each round.  
Moreover, each client in the federated system can leverage off-the-shelf acceleration frameworks (e.g., vLLM \cite{kwon2023efficient}) to enhance inference efficiency, thereby enabling our approach to scale to larger models. For instance, under the same execution environment, employing vLLM for data synthesis introduces an additional latency of approximately 65 seconds for LlaMa2-7B, while LlaMa2-13B requires around 122 seconds.
Importantly, \ourmethod significantly enhances training performance, with improvements ranging from 7\% to 13\%, and reduces data leakage risk by approximately 20\%. \textbf{Given the substantial benefits, we argue that these additional computational costs can be deemed acceptable.}

%% file: results/datasets.tex
\begin{table}
    \centering
    \caption{The data statistics and federated setting in our experiment.}\label{tab: selection}\label{tab: data_fl}
    \small
    \begin{tabular}{l|cc|ccc}
        \toprule
        & \multicolumn{2}{c|}{Data Statistics} & \multicolumn{3}{c}{Federated Setting} \\
        Datasets & |Train| & |Test|  & Scenario & Rounds & |Clients|  \\ \midrule
        \alpaca & 1000 & 805 & Cross-device & 30 & 50 \\  \midrule 
        \alpacare & 500 & 216 & Cross-silo & 10 & 10 \\ \midrule
        \medalpaca & 1000 & 400 & Cross-device & 30 & 50 \\ 
        \bottomrule
    \end{tabular}
\end{table}

%% file: results/utility.tex
\begin{table*}[t]
    \caption{The performance of \ourmethod and other contenders on federated few-shot instruction tuning. \ourmethod surpasses all federated algorithms and closely approaches the skyline algorithm \texttt{Centralized} performance.}
    \label{tab: performace}
    \centering
    \resizebox{\linewidth}{!}{
    \begin{tabular}{l|ccc|ccc|ccc|ccc}
    \toprule
     & \multicolumn{3}{c|}{\alpaca} & \multicolumn{3}{c|}{\alpacare} & \multicolumn{3}{c|}{\medalpaca} & \multicolumn{3}{c}{Avg.} \\
     Methods & Win ($\uparrow$) & Tie ($\uparrow$) & Lose ($\downarrow$) & Win ($\uparrow$) & Tie ($\uparrow$) & Lose ($\downarrow$) & Win ($\uparrow$) & Tie ($\uparrow$) & Lose ($\downarrow$) & Win ($\uparrow$) & Tie ($\uparrow$) & Lose ($\downarrow$) \\ \midrule
     \cenit & 23.0 & 34.9 & 42.1 & 26.1 & 59.8 & 14.1 & 32.4 & 16.2 & 51.4 & 27.2 & 37.0 & 35.9     \\
     \midrule
     \fedavg    & 11.3 & 29.2 & 59.5 & 20.8 & \textbf{58.5} & 20.7 & 29.2 & 12.8 & 58.0 & 20.4 & 33.5 & 46.1 \\
     \fedprox   & 12.3 & 30.4 & 57.3 & 22.5 & 58.0 & 19.5 & 28.5 & 13.2 & 58.3 & 21.1 & 33.9 & 45.0 \\
     \scaffold  & 11.4 & 31.6 & 57.0 & 22.0 & \textbf{58.5} & 19.5 & 29.0 & 10.5 & 60.5 & 20.8 & 33.5 & 45.7 \\ 
     \fedopt   & 10.2 & 29.1 & 60.7 & 18.5 & 53.0 & 28.5 & 26.0 & 10.1 & 63.9 & 18.2 & 30.7 & 51.0 \\
     \midrule
     \fewfed & 15.3 & 29.6 & 55.1 & 23.1 & 57.2 & 19.7 & 28.7 & 13.1 & 58.2 & 22.4 & 33.3 & 44.3 \\
     \ourmethod & \textbf{22.3} & \textbf{34.0} & \textbf{43.7} & \textbf{25.5} & \textbf{58.5} & \textbf{16.0} & \textbf{30.8} & \textbf{14.9} & \textbf{54.3} & \textbf{26.2} & \textbf{35.8} & \textbf{38.0}
     \\ \bottomrule
    \end{tabular}
    }
    
\end{table*}

%% file: 5_Conclusion.tex
This paper proposes a \textit{novel} federated algorithm, \ourmethod, that leverages LLMs' in-context learning capability to generate task-specific synthetic data, improving federated few-shot performance and against training data extraction attacks. 
To reduce noise in the synthesized data, we select high-quality synthesized data using instruction-following scores and propose parameter isolation training to reduce the effect of noisy data. 
Inspired by experimental findings that federated aggregation can reduce privacy leakage, we implement local aggregation sharing, which mixes public parameters and private parameters before uploading to the server.
Through extensive experiments on three open-source datasets, we demonstrate the effectiveness of \ourmethod in enhancing federated few-shot performance while defending against data extraction attacks. 
Our contributions pave the way for more robust and privacy-preserving FL approaches, particularly in privacy-sensitive domains where data scarcity and privacy concerns are paramount.

%% file: 6_acknowledgments.tex
This work was supported by the National Natural Science Foundation of China (62472125), Guangdong Basic and Applied Basic Research Foundation (2025A1515011258), Key Technologies R\&D Program of Guangdong Province (2026B0909060001) and Shenzhen Science and Technology Programs (GXWD20231128102922001, ZDCY20250901111705007, ZDSYS20230626091203008).

%% file: 7_Appendix.tex
\newpage
\appendix

\section{Dataset and Training Details}\label{sec: app_datasets}
Our experiments use three open-source instruction datasets, including one open domain \alpaca~\cite{peng2023instruction}, and two medical domains \alpacare~\cite{zhang2023alpacare} and \medalpaca~\cite{han2023medalpaca}. 
\alpaca is the popular instruction-following dataset with 52k examples generated by GPT-4 using Alpaca prompts for fine-tuning LLMs. 
\alpacare is a diverse, machine-generated medical instruction-following dataset with 52k instances, using GPT-4 and ChatGPT with a high-quality expert-curated seed set. 
\medalpaca is an innovative dataset consisting of over 514k entries, specifically crafted to fine-tune LLMs for effective medical applications. We follow \citet{zhao2024enhancing} and adopt question-answering task~\cite{jin2020disease}. 
These medical datasets are sensitive, involving medical consultations and clinical narratives, thereby serving as realistic proxies for privacy-restricted environments governed by regulations such as HIPAA and GDPR.
We downsample these datasets to construct federated few-shot scenarios. Considering the critical role of instruction diversity, we follow previous work using the KMeans algorithm for selection. First, we encoded each instance using the longformer~\cite{beltagy2020longformer}, followed by 100 dataset clustering, and finally, selecting 10 (\alpacare and \alpaca) or 5 (\medalpaca) samples from each cluster. 

Our experiments use an open-form assessment for model utility evaluation. For \alpaca, we use the well-known dataset called AlpacaEval~\cite{alpaca_eval}, which compares the model’s generated responses to Davinci-003 responses. For \alpacare, we follow \citet{zhang2023alpacare} and use MedInstructTest, a dataset their clinicians created, including 216 medical instructions. To be consistent with the previous test method, in \medalpaca, we randomly selected 400 entries from the remaining data as the test dataset, ensuring no overlap with the training dataset.

We run a hyperparameter sweep for each dataset and tuning method to make a fair and reasonable comparison. 
Especially, the learning rate is selected from {3e-4, 2e-3}.
Following ~\cite{ye2024openfedllm}, we apply a cosine learning rate schedule according to the round index. 
For all datasets, we use the Alpaca template to format the instructions. 
We quantize the backbone model using int8 and enable bfloat16 to improve computational efficiency. 
The code will be made publicly available on GitHub after the review process.
All experiments are conducted on a server with 2 Nvidia A100 GPUs with 40GB RAM each.

\section{Prompts}\label{sec: app_prompts}
\begin{figure*}[t]
\centering
\includegraphics[trim={0cm 10cm 10cm 0cm}, clip, width=\linewidth]{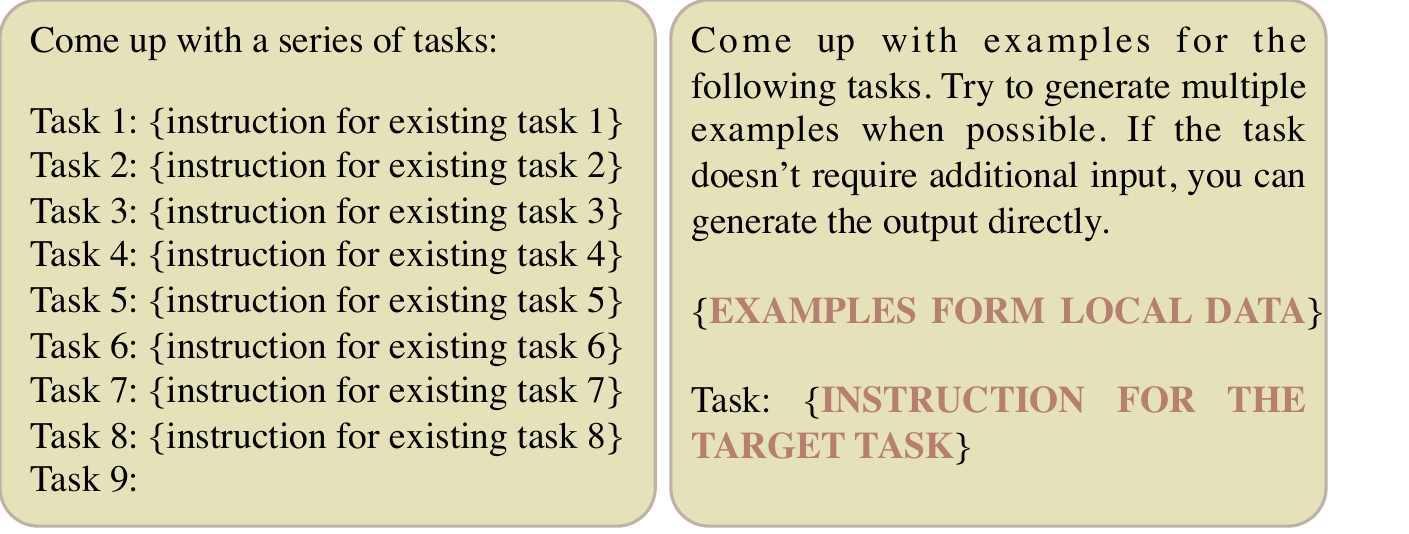}
\caption{The prompts used in our synthetic data generation. (a) Prompt used for generating new instructions. We randomly sample eight instructions from local data for in-context demonstration. The model is allowed to generate instructions for the new instruction. (b) Prompts used for input and output generation given the new instruction. We prompt the model with randomly sampled examples followed by the new instruction for the response generation. } 
\label{fig: prompt_sg} 
\end{figure*}

\paragraph{Synthetic data generation prompts.} Fig.~\ref{fig: prompt_sg} illustrates the prompts used in synthetic data generation, adapted from \citet{wang2022self}. Our approach utilizes Fig.~\ref{fig: prompt_sg} (a) to generate new INSTRUCTIONS. We randomly sample eight instructions from local data for in-context demonstration. The model is allowed to generate instructions for the new instruction. Fig.~\ref{fig: prompt_sg} (b) shows the prompt to generate corresponding RESPONSES. We prompt the model with randomly sampled examples followed by the new instruction for the response generation.

\begin{figure*}[t]
\centering
\includegraphics[trim={0cm 0cm 14.7cm 0cm}, clip, width=\linewidth]{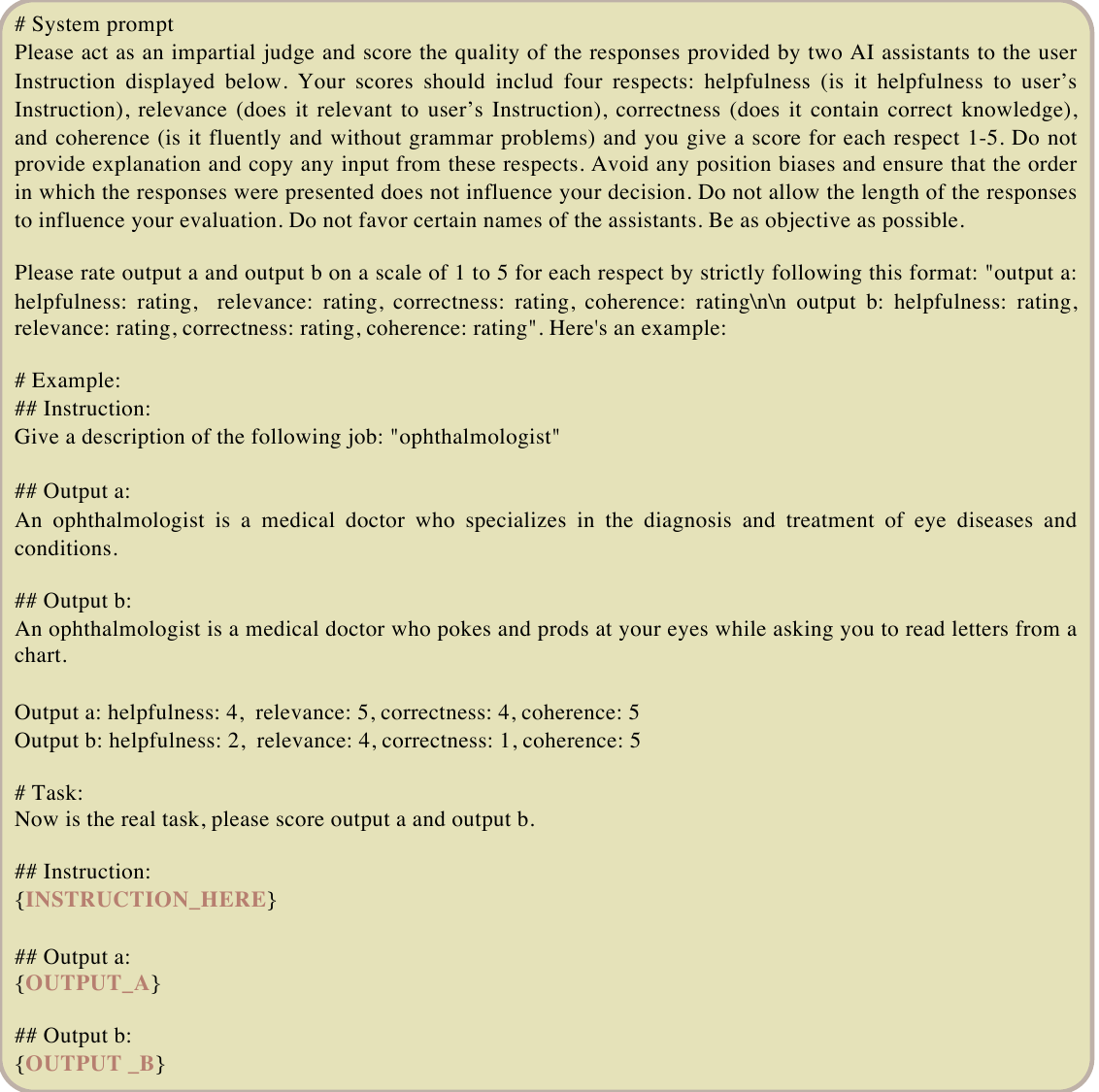}
\caption{GPT4-as-a-Judge prompt for evaluating the outputs of \ourmethod and baseline methods.} 
\label{fig: eval_prompt} 
\end{figure*}

\paragraph{GPT4-as-a-Judge prompt.} Fig.~\ref{fig: eval_prompt} illustrates the GPT-4-as-a-Judge prompt used in our experimental evaluation, adapted from the LLMs evaluation prompt introduced by \citet{zheng2024judging}. To enhance evaluation quality, GPT-4 assesses the output of both methods across four respects: helpfulness, relevance, correctness, and coherence. The average score across these respects is the method's overall score, determining win, tie, and loss ratios. We conduct two rounds of testing with alternating positional substitutions to mitigate potential positional biases in GPT-4~\cite{wang2023large}.

\section{Synthetic Data Analysis}\label{app_quality}
\input{results/quality}

We explore the quality of the synthetic data generated. 
Following \citet{wang2022self}, we assess whether the data in the synthetic data generation step is correct for each instance regarding instructions, instance inputs, and instance outputs. 
The review results are shown in Table~\ref{tab: quality_review}. Although the synthetic data contain errors and noise, most are still correct, which can improve the local model training. This result also reflects the necessity of parameter isolation training and high-quality data filtering.
Additionally, we use the generated synthetic data to directly train small LLMs, such as TinyLlama-1.1B~\cite{zhang2024tinyllama}, to validate the data quality. 
We use the synthetic data generated on \alpaca and test on the HuggingFace's OpenLLM evaluation benchmark~\cite{gao2021framework}, as shown in Table~\ref{tab: quality_train}. The experiments demonstrate that the synthetic data generation mechanism we designed can produce helpful instruction-tuning data.

\section{Theoretical analysis for \ourmethod}\label{sec: app_theo}
Benefiting from the local aggregation sharing mechanism, \ourmethod effectively mitigates data extraction attacks while achieving a more favorable trade-off between performance and privacy protection. To gain a deeper understanding of this privacy-preserving property, we first provide a theoretical analysis demonstrating that local aggregation sharing can approximate differential privacy without requiring explicit noise injection or gradient clipping. Furthermore, we establish that during the federated aggregation phase, this mechanism substantially increases the computational complexity of potential attacks, thereby reinforcing the overall privacy guaranties.

\subsection{Differential privacy approximation of the local aggregation sharing in \ourmethod}
\subsubsection{Definitions and Background}
We first provide the definitions and background of local aggregation sharing,  adjacent datasets, global sensitivity, and local differential privacy.

\textbf{Local Aggregation Sharing.} Let the local model trained on a private dataset $\mathcal{D}$ produce parameters $\mathcal{W}_l$, and the model trained on synthetic data (containing no private information) produce parameters $\mathcal{W}_l$. The local aggregation sharing mechanism combines these two components as: $\mathcal{W}^{a} = \beta*\mathcal{W}^{l} + (1-\beta) * \mathcal{W}^{s}$. $\beta$ is a fixed mixing coefficient. A smaller $\beta$ implies that a greater proportion of the shared parameters arises from non-private, synthetic data, thereby attenuating the privacy exposure.

\textbf{Adjacent Datasets.} Two datasets $\mathcal{D}$ and $\mathcal{D}^{\prime}$ are adjacent if they differ by at most one sample; formally, $\mathcal{D}^{\prime}$ can be obtained from $\mathcal{D}$ by adding, removing, or replacing a single record~\cite{dwork2014algorithmic}. This notion captures the smallest perturbation in an individual client’s data.

\textbf{Global Sensitivity.} For a deterministic function $f: \mathcal{D} \rightarrow \mathbf{R}^d$ (e.g., a training algorithm returning model parameters), its global sensitivity is:
\begin{equation}
\Delta f=\max _{\mathcal{D} \sim \mathcal{D}^{\prime}}\left\|f(\mathcal{D})-f\left(\mathcal{D}^{\prime}\right)\right\|,
\end{equation}
where $\mathcal{D} \sim \mathcal{D}^{\prime}$. $\Delta f$ quantifies the maximum change in the output that can result from modifying a single training record.

\textbf{Local Differential Privacy.} A randomized mechanism $\mathcal{M}$ satisfies $(\epsilon, \delta)$\textit{-local differential privacy} (LDP), for any adjacent datasets $\mathcal{D}$, $\mathcal{D}^{\prime}$ and measurable subset $S$:
\begin{equation}
\operatorname{Pr}[\mathcal{M}(D) \in S] \leq e^\epsilon \operatorname{Pr}\left[\mathcal{M}\left(D^{\prime}\right) \in S\right]+\delta.
\end{equation}

When $\delta=0$, the guaranty is pure $\epsilon$-LDP~\cite{dwork2014algorithmic}. This condition ensures that the inclusion or exclusion of any single record changes the output distribution only negligibly, limiting an adversary’s ability to infer the presence of that record~\cite{kasiviswanathan2011can,dwork2014algorithmic}.

\subsubsection{Sensitivity reduction by local aggregation sharing}
\begin{lemma}
Let $f(\mathcal{D})=\mathcal{W}^{l}_{\mathcal{D}}$ denote the parameters trained on the private data $\mathcal{D}$, and define the local aggregation parameters:
\begin{equation}
g(\mathcal{D})=W^a_{\mathcal{D}}=\beta W^l_{\mathcal{D}}+(1-\beta) W^s,
\end{equation}
where $\mathcal{W}^s$ is trained solely on synthetic data and is therefore independent of $\mathcal{D}$. Then the global sensitivity of $g(\mathcal{D})$ equals $\beta$ times that of $f(\mathcal{D})$:
\begin{equation}
\Delta \mathcal{W}^a = \beta \Delta \mathcal{W}^l.
\end{equation}
\end{lemma}

\begin{proof}
For any adjacent datasets $\mathcal{D}$, $\mathcal{D}^{\prime}$, we have:
\begin{equation}
\mathcal{W}^a_{\mathcal{D}}-\mathcal{W}^a_{\mathcal{D}^{\prime}}=\beta\left[\mathcal{W}^l_{\mathcal{D}}-\mathcal{W}^l_{\mathcal{D}^{\prime}}\right]+(1-\beta)\left[\mathcal{W}^s_{\mathcal{D}}-\mathcal{W}^s_{\mathcal{D}^{\prime}}\right]=\beta\left[\mathcal{W}^l_{\mathcal{D}}-\mathcal{W}^l_{\mathcal{D}^{\prime}}\right],
\end{equation}
since $\mathcal{W}^s_{\mathcal{D}} = \mathcal{W}^s_{\mathcal{D}^{\prime}}$. Taking norms and maximizing over all adjacent pairs gives:
\begin{equation}
\Delta \mathcal{W}^a=\max _{\mathcal{D} \sim \mathcal{D}^{\prime}}\left\|\mathcal{W}^a_\mathcal{D}-\mathcal{W}^a_{\mathcal{D}^{\prime}}\right\|=\beta \max_{\mathcal{D} \sim \mathcal{D}^{\prime}}\left\|\mathcal{W}^l_\mathcal{D}-\mathcal{W}^l_{\mathcal{D}^{\prime}}\right\|=\beta \Delta \mathcal{W}^l .
\end{equation}
Thus, the contribution of any single data sample to the uploaded parameter vector is scaled down by a factor $\beta$. When $\beta < 1$, each coordinate’s sensitivity decreases proportionally, reducing the dependence of the public parameters on private data.
\end{proof}

\subsubsection{Approximate Differential Privacy Guaranty}
\begin{theorem}
The local aggregation sharing mechanism approximately satisfies $(\epsilon, \delta)$\textit{-LDP} even without explicit noise injection. For any adjacent datasets $\mathcal{D}$, $\mathcal{D}^{\prime}$ the distributions of $\mathcal{W}^a_{\mathcal{D}}$ and $\mathcal{W}^a_{\mathcal{D}^{\prime}}$ are statistically close, showing small total-variation distance, and hence meet an approximate LDP criterion.
\end{theorem}

\begin{proof}
In practice, randomness arises inherently from stochastic gradient descent, random initialization, and dropout, making $\mathcal{W}^l_{\mathcal{D}}$ a random variable. Consequently, $W^a_{\mathcal{D}}=\beta W^l_{\mathcal{D}}+(1-\beta) W^s$ is also random. Let $P_\mathcal{D}$ and $P_{\mathcal{D}^{\prime}}$ denote the probability distributions of $W^a_{\mathcal{D}}$ and $W^a_{\mathcal{D}^{\prime}}$. The total-variation (TV) distance is
\begin{equation}
\Delta_{\mathrm{TV}}\left(P_\mathcal{D}, P_{\mathcal{D}^{\prime}}\right)=\sup _S\left|\operatorname{Pr}\left[\mathcal{W}^a_{\mathcal{D}} \in S\right]-\operatorname{Pr}\left[\mathcal{W}^a_{\mathcal{D}^{\prime}} \in S\right]\right| .
\end{equation}
By Lemma 1, the two random outputs differ by at most $\|\mathcal{W}^a_{\mathcal{D}}-\mathcal{W}^a_{\mathcal{D}^{\prime}} \leq \beta \Delta \mathcal{W}^l\|$. Assuming $P_\mathcal{D}$ admits a Lipschitz-continuous density (as in SGD-trained models) \cite{abadi2016deep}, such a small shift alters only a tiny portion of probability mass. Therefore, there exists a small constant $\delta = \mathcal{O}(\beta \Delta \mathcal{W}^l)$ such that:
\begin{equation}
\Delta_{\mathrm{TV}}\left(P_\mathcal{D}, P_{\mathcal{D}^{\prime}}\right) \leq \delta.
\end{equation}
By Pinsker’s inequality~\cite{dwork2014algorithmic}, bounded TV distance implies approximate $(\epsilon, \delta)$\textit{-LDP}: for sufficiently small $\beta$, the two output distributions are virtually indistinguishable to any observer.
\end{proof}

\subsubsection{Summary} The local aggregation sharing mechanism in \ourmethod inherently attenuates per-sample sensitivity, thereby achieving approximate differential privacy without explicit random noise. Formally,

\begin{equation}
\Delta W^a=\beta \Delta W^l, \Delta_{\mathrm{TV}}\left(P_\mathcal{D}, P_{\mathcal{D}^{\prime}}\right) \leq \mathcal{O}(\beta \Delta \mathcal{W}^l).
\end{equation}

Thus, reducing $\beta$ directly strengthens privacy.
Even without traditional DP noise, this mechanism provides a mathematically grounded privacy-attenuation effect, limiting adversarial information gain and offering provable resistance to training data reconstruction attacks.

\subsection{Federated aggregation and K-Vector Subset Sum}
In \ourmethod, the mechanism of local aggregation sharing further strengthens privacy protection during the federated aggregation phase. When multiple clients participate, the server receives mixed parameters of the form:
\begin{equation}
    \mathcal{W}^{a}_{k} = \beta \mathcal{W}^{l}_{k} + (1-\beta)\mathcal{W}^{s}_{k},
\end{equation}
where $k$ denotes the $k$-th client, and updates the global model $\mathcal{W}^{g}$ by performing weighted averaging or summation across these mixed parameters. Crucially, each uploaded $\mathcal{W}^{a}$ represents a linear combination of private and shared parameters, thereby constructing a “multi-vector subset-sum” obfuscation in the aggregation outcome. Specifically, the global update can be expressed as:
\begin{equation}
    \mathcal{W}^g=\sum_k p_k \mathcal{W}^a_{k}=\beta \sum_k p_k \mathcal{W}^l_{k}+(1-\beta) \sum_k p_k \mathcal{W}^s_{k},
\end{equation}
where $p_k$ denotes the aggregation weight of the $k$-th client. From the perspective of the server or a potential adversary, only the mixed vectors and their aggregated result are observable, making it computationally intractable to disentangle individual $\mathcal{W}^l_{k}$ and $\mathcal{W}^s_{k}$.

Attempting to recover a client’s pure local parameters $\mathcal{W}^l_{k}$ from the mixed representation amounts to solving a high-dimensional demixing problem: decomposing the known $\mathcal{W}^a_{k}$ into $\mathcal{W}^l_{k}$ and $\mathcal{W}^s_{k}$. Since $\mathcal{W}^s_{k}$ is trained from synthetic data unique to each client and unknown to the adversary, it effectively acts as a client-specific noise mask. Even with knowledge of the mixing formula and the coefficient $\beta$ the absence of prior information about $\mathcal{W}^s_{k}$ prevents tractable inference. The presence of multiple clients does not simplify the problem: the adversary must simultaneously solve for $k$ unrelated mixed vectors, recovering both $\mathcal{W}^a_{k}$ and $\mathcal{W}^l_{k}$ for each client.

This challenge closely parallels the \textit{k-Vector Subset Sum} problem~\cite{abboud2014losing}, a well-known computationally intractable task. As prior work such as TextHide\cite{huang2020texthide} has shown, mixing sensitive data before disclosure forces adversaries to confront an inverse subset-sum problem, whose complexity is exponential in the worst case. Indeed, the k-Vector Subset Sum problem has been proven to be NP-complete, implying that an adversary would require prohibitive supercomputing resources to exhaustively enumerate all possible combinations. Consequently, reconstructing individual clients’ private parameters $\mathcal{W}^l_{k}$ within reasonable time is practically infeasible.

%% file: results/quality.tex
\begin{table}
    \centering
    \caption{Data quality review of the generated synthetic data.}\label{tab: quality_review}
    \begin{tabular}{c|ccc}
        \toprule
        Quality Review Question & \alpaca & \alpacare & \medalpaca \\ \midrule
        \makecell{Does the instruction \\ describe a valid task?} & 91\%  & 88\% & 85\% \\ \midrule
        \makecell{Is the response appropriate \\ for the instruction and input?} & 52\%  & 59\% & 57\% \\ \midrule
        \makecell{Is the correct format?} & 56\%  & 66\% & 53\%  \\
        \bottomrule
    \end{tabular}
    
\end{table}

\begin{table}
    \centering
    \caption{The performance of TinyLlama-1.1B on HuggingFace OpenLLM Benchmark using synthetic data of \alpaca. \texttt{Pre-train} refers to the backbone model without instruction tuning.}\label{tab: quality_train}
    \begin{tabular}{l|ccccc|c}
        \toprule
        & Winogrande & ARC & Hellaswag & TruthfulQA & MMLU & Avg. \\
        \midrule
        Pre-train & 57.5  & 48.1  & 44.8  & 37.2  & 24.2  & 42.4 \\ \midrule
        Ours & 60.1  & 50.5  & 59.0  & 39.4  & 25.6 & 46.9 \\
        \bottomrule
    \end{tabular}
\end{table}